\documentclass[pre,twocolumn,showpacs]{revtex4}
\usepackage{epsfig}
\usepackage{amssymb}
\begin{document}

\title{Water wave collapses over quasi-one-dimensional 
non-uniformly periodic bed profiles} 
\author{V. P. Ruban}
\email{ruban@itp.ac.ru}
\affiliation{Landau Institute for Theoretical Physics,
2 Kosygin Street, 119334 Moscow, Russia} 

\date{\today}

\begin{abstract}
Nonlinear water waves interacting with quasi-one-dimensional, 
non-uniformly periodic 
bed profiles are studied numerically in the deep-water regime with the help of
approximate equations for envelopes of the forward and backward waves. 
Spontaneous formation of localized two-dimensional wave structures is observed 
in the numerical experiments, which looks essentially as a wave collapse.
\end{abstract}

\pacs{47.15.K-, 47.35.Bb, 47.35.Lf}

%47.15.K- Inviscid laminar flows
%47.35.Bb Gravity waves
%47.35.Lf Wave-structure interactions

\maketitle

Nonlinear waves of different nature, when propagating in spatially periodic 
media, are known to introduce various interesting phenomena. 
In particular, the so called gap solitons (GSs) can be mentioned, 
which are self-localized waves existing due to the presence of a Bragg gap 
in the spectrum of linear waves, and, on the other hand, due to nonlinear 
interactions between components of the wave field. Gap solitons are studied 
mainly in the nonlinear optics (see, e.g., 
Refs.\cite{CM1987,AW1989,CJ1989,ESdSKS1996,
PPLM1997,BPZ1998,RCT1998,CTA2000,IdS2000,CT2001,CMMNSW2008}), and 
in the theory of Bose-Einstein condensation \cite{EC2003,PSK2004,MKTK2006}. 
However, recently it has been suggested that water-wave GSs are possible too, 
over  periodic bottom boundary \cite{R2008PRE-2,R2008PRE-3}. 
In the cited works \cite{R2008PRE-2,R2008PRE-3}, planar potential flows 
with one-dimensional (1D) free boundary were studied both
analytically --- with the help of an approximate model possessing relatively 
simple particular solutions, and numerically ---
exact equations of motion for an ideal fluid with a free surface 
over a non-uniform bed were simulated, in terms of the so called conformal 
variables (see Ref.\cite{R2004PRE}). 
For two horizontal dimensions, the question about possibility of 
GSs or some other coherent water-wave structures over periodic bed profiles 
is still open. The present work is a step to study this problem.
More specifically, we suggest here approximate equations of motion for a
two-dimensional (2D) free surface over a quasi-one-dimensional, locally 
periodic non-uniform bed. Then we present some numerical results
where spontaneous formation of localized nonlinear structures is clearly
seen, and the process looks as a kind of wave collapse. 
It should be emphasized that the Bragg interaction between the forward wave 
and the backward wave plays the key role in this process, 
since the wave dynamics over a flat bottom is rather different, 
and the extreme waves there are not so high as 
they are with the Bragg interaction. It is quite possible that effects
considered in the present work have in some cases relation to the 
extensively discussed phenomenon of rogue waves 
\cite{Kharif-Pelinovsky,RogueWaves2006}, 
namely at seas with a non-uniform depth $\sim 50 - 100$ m.

Analytical and numerical results of Ref.\cite{R2008PRE-3} imply that
the most suitable asymptotic regime for observation of nonlinear 
Bragg structures of the GS type in water-wave systems 
is the relatively short-wave (or deep-water) regime, 
when an effective water depth $h_0$ and the 
main Bragg-resonant wave number $\kappa$ form a small parameter 
$\varepsilon \equiv\exp(-2\kappa h_0)\sim 0.01$. 
It should be noted here that the opposite long-wave asymptotic regime is 
very different, and typical nonlinear structures at shallow water are KdV-type 
solitons (see, Refs.\cite{MeiLi2004,GKN2007}).
In the considered deep-water regime, one can use the 
fourth-order Hamiltonian functional 
${\cal H}\{\eta,\psi\}$ for weakly-nonlinear gravity water waves in the 
approximate form (see Ref.\cite{R2008PRE-3}),
\begin{eqnarray}
{\cal H}&\approx&\frac{1}{2}\int\left\{\psi\hat K\psi +g\eta^2
+\eta \left[(\nabla \psi)^2-(\hat k\psi)^2\right]\right\} d^2{\bf r}
\nonumber\\
&+&\frac{1}{2}\int\left[
\psi\hat k\eta \hat k\eta\hat k\psi 
+\eta^2(\hat k\psi)\nabla^2\psi \right] d^2{\bf r},\label{H_4}
\label{H_approx}
\end{eqnarray}
where ${\bf r}=(x,y)$ is the position in the horizontal plane, $g$ is the
gravity acceleration, the canonical variables $\eta({\bf r},t)$ and 
$\psi({\bf r},t)$ are the vertical coordinate of the free surface 
and the boundary value of the velocity potential respectively, 
the linear operator $\hat k=(\hat k_x^2+\hat k_y^2)^{1/2}$ is  
diagonal in the Fourier representation, and $\hat K=\hat k+\hat B$ is
a linear operator connecting the velocity potential $\varphi(x,y, z=0)$ 
at the unperturbed free surface and the quantity $\partial_z\varphi(x,y, z=0)$,
in the presence of an inhomogeneous bed. The operator $\hat B$ depends on a 
given bed profile in a complicated manner (see 
Refs.\cite{RA_Smith1998,R2008PRE-3}, where
some expansions of this operator are discussed), 
but it is definitely ``small'' in the deep-water limit, 
$\hat B\sim \varepsilon$, and that is why we keep $\hat B$ only in the 
quadratic part of the Hamiltonian, while in all higher-order terms we write 
$\hat k$ instead of $\hat K$. 

Canonical equations of motion, corresponding to Hamiltonian 
(\ref{H_approx}), are still too complicated to be treated analytically,
and very time-demanding when being solved numerically. Therefore we shall
consider here a simplified model which can be derived from 
(\ref{H_approx}) in some limit.
To do this, let us recall the well known fact that a suitable weakly 
nonlinear canonical transformation, 
\begin{equation}
b_{\bf k}=\frac{\sqrt{g}\eta_{\bf k}+i\sqrt{|{\bf k}|}\psi_{\bf k}}
{\sqrt{2\omega_{\bf k}}}+\mbox{ nonlinear terms},
\end{equation}
(where $\omega_{\bf k}=(g|{\bf k}|)^{1/2}$ is a linear dispersion relation 
in the absence of the bottom boundary) can exclude the third-order terms from 
the deep-water Hamiltonian, as well as the non-resonant wave interactions
$0\leftrightarrow 4$ and  $1\leftrightarrow 3$ (see Refs.\cite{K1994,Z1999}). 
Accordingly, the shape of the free surface $\eta(x,y,t)$ is given by the
formula below:
\begin{equation}
\eta({\bf r},t)=\mbox{Re}\int
\left(\frac{2\omega_{\bf k}}{g}\right)^{1/2} 
b_{\bf k}(t)e^{i{\bf k r}}d{\bf k}/(2\pi) +\cdots.
\end{equation}
In terms of the new normal complex variables $b_{\bf k}(t)$, the wave dynamics 
in deep-water regime is described by the following integral equation,
\begin{eqnarray}
i\dot b_{\bf k}&\approx&\omega_{\bf k} b_{\bf k}
+\hat L b_{\bf k}+\dots
+\frac{1}{2}\int T({\bf k},{\bf k}_2;{\bf k}_3,{\bf k}_4)
b^*_{{\bf k}_2}b_{{\bf k}_3}b_{{\bf k}_4}\nonumber\\
&&\qquad\qquad\times\delta({\bf k}+{\bf k}_2-{\bf k}_3-{\bf k}_4)
\,d{\bf k}_2\,d{\bf k}_3\,d{\bf k}_4,
\label{b_k_equation}
\end{eqnarray}
where $\hat L$ is a ``small'' linear non-diagonal operator related to $\hat B$,
and $T({\bf k}_1,{\bf k}_2;{\bf k}_3,{\bf k}_4)$ is a known continuous function
(but the corresponding explicit expression is rather complicated, 
see Refs.\cite{K1994,Z1999}). For our purposes it is sufficient to know that
in purely one-dimensional case 
\begin{equation}\label{T1212}
T(k_1,k_2; k_1,k_2)\propto |k_1||k_2|(|k_1+k_2|-|k_1-k_2|),
\end{equation}
and thus $T(k,-k; k,-k)=-T(k,k; k,k)<0$.

Let us assume for the moment that a bed profile is strictly $x$-periodic,
with the period $\Lambda=\pi/\kappa$,  and a 1D wave spectrum is concentrated 
near the Bragg-resonant wave vectors $\pm {\bf k}_0=\pm\kappa{\bf e}_x$.
Now we introduce two slowly-$x$-dependent functions, 
the envelopes $A_\pm(x,t)$ of the forward- and backward-propagating waves, 
$$
\eta(x,t)=\mbox{Re}[A_+\exp(i\kappa x-i\tilde\omega t)
+A_-\exp(-i\kappa x-i\tilde\omega t)],
$$
where $\tilde\omega=(g\kappa)^{1/2}$ (the corresponding deep-water wave period 
is $T_0=2\pi/\tilde\omega$).
Then we take into account only the main-order
components of the integral operator $\hat L$: 
$$
L_{{\bf k}_0,{\bf k}_0}=L_{-{\bf k}_0,-{\bf k}_0}=-\tilde\omega\varepsilon,
\quad 
L_{{\bf k}_0,-{\bf k}_0}=L^*_{-{\bf k}_0,{\bf k}_0}= \tilde\omega\Delta,
$$
where $\varepsilon$ is a real positive number, which should be identified as
$\varepsilon=\exp(-2\kappa h_0)$, for some efficient depth $h_0$,
since the finite-depth dispersion relation is
$\omega=
[g\kappa\tanh(h_0\kappa)]^{1/2}\approx\tilde\omega[1-\exp(-2\kappa h_0)]$;
a complex parameter $\Delta=|\Delta|\exp(i\phi_0)$ 
bears information about the width of the main Bragg gap (namely,
the edges of the frequency gap are at
$\omega_{1,2}\approx \tilde\omega(1-\varepsilon\mp|\Delta|$), 
and about the phase $\phi_0$ of 
the main Fourier harmonics of the bed undulation in some suitable
representation (see Ref.\cite{R2008PRE-3} for details).
It is worth mentioning that for all 1D bed profiles considered in the previous 
studies \cite{R2008PRE-2,R2008PRE-3}, the inequality 
$|\Delta|<\varepsilon$ holds ($|\Delta|$ approaches $\varepsilon$ from the 
below if the bottom boundary consists of very narrow barriers).

As a result of the standard procedure, we obtain the system of two coupled 
approximate equations  for the wave envelopes $A_\pm(x,t)$ 
(compare to Ref.\cite{R2008PRE-3}),
\begin{eqnarray}
\label{Ap1D}
&&i\Big(\frac{\partial_t}{\tilde\omega}+\frac{\partial_x}{2\kappa}
+\cdots\Big)A_+=\Delta\, A_- -\varepsilon A_+
\nonumber\\ 
&&\qquad\qquad+\frac{\kappa^2}{2}\left(|A_+|^2-2|A_-|^2\right)A_+,
\label{A_plus_eq}
\\
\label{Am1D}
&&i\Big(\frac{\partial_t}{\tilde\omega}-\frac{\partial_x}{2\kappa}
+\cdots\Big)A_-=\Delta^*\, A_+ -\varepsilon A_-
\nonumber\\ 
&&\qquad\qquad +\frac{\kappa^2}{2}\left(|A_-|^2-2|A_+|^2\right)A_-,
\label{A_minus_eq}
\end{eqnarray}
where the dots mean the second- and higher-order $x$-derivatives.
The above equations describe weakly nonlinear 1D water-wave GSs 
quite well: known solitary-wave solutions of this system
were compared in Ref.\cite{R2008PRE-3} to numerical simulations of 
the exact equations of motion for an ideal fluid with a free surface, 
and a reasonable correspondence was found.

Now it is clear how to generalize the above system to the 2D case:
we have just to add $\partial_y$-dependent dispersive terms to the 
left hand sides, which appear from expansions of the operators 
$[-\kappa^{-1/2}(( -i\partial_x\pm \kappa)^2-\partial_y^2)^{1/4}+1]$.
Moreover, we shall consider here the case when $\varepsilon$ and $\Delta$ 
are no longer constants, but they are some functions slowly depending on 
the horizontal coordinates $x$ and $y$. This assumption makes our model 
more realistic and rich, since there is no perfectly periodic 
bed structures in the nature, while roughly periodic bars are quite often.
Then the following natural 
generalization of Eqs.(\ref{Ap1D}-\ref{Am1D}) can be suggested,
\begin{eqnarray}
&&\Big(\frac{i\partial_t}{\tilde\omega}+
\frac{i\partial_x}{2\kappa}-\frac{\partial_x^2}{8\kappa^2}
+\frac{\partial_y^2}{4\kappa^2}+\cdots\Big)A_+ =-\varepsilon(x,y)A_+
\nonumber\\
&&\qquad+\Delta(x,y) A_- 
 +\frac{\kappa^2}{2}\left(|A_+|^2-2|A_-|^2\right)A_+,
\label{Aplus_2D_model_1Dbed} 
\\
&&\Big(\frac{i\partial_t}{\tilde\omega}-
\frac{i\partial_x}{2\kappa}-\frac{\partial_x^2}{8\kappa^2}
+\frac{\partial_y^2}{4\kappa^2}+\cdots\Big)A_- = -\varepsilon(x,y)A_-
\nonumber\\
&&\qquad+\Delta^*(x,y) A_+ 
 +\frac{\kappa^2}{2}\left(|A_-|^2-2|A_+|^2\right)A_-,
\label{Aminus_2D_model_1Dbed} 
\end{eqnarray}
where the dots mean omitted third- and higher-order partial derivatives.
We keep here the second-order dispersive terms because they are 
definitely important in the dynamics of relatively short wave groups.
The linear and nonlinear terms in the right hand sides
are of the same order of magnitude at $\kappa |A_\pm|\sim 0.1$ and 
$\varepsilon\sim|\Delta|\sim 0.01$ (let us note that water waves become 
strongly nonlinear if $\kappa |A_\pm|\gtrapprox 0.3$).
The above equations have the standard Hamiltonian structure, 
$i\tilde\omega^{-1}\partial_tA_\pm=\delta\tilde{\cal H}/\delta A^*_\pm$, with
\begin{eqnarray}
\tilde{\cal H}&=&
\int [A^*_+\hat D_+ A_+ + A^*_-\hat D_- A_-]d^2{\bf r}
\nonumber\\
&+&\int [\Delta A^*_+ A_- + \Delta^* A^*_-A_+ 
-\varepsilon(A^*_+ A_+ + A^*_- A_-)]d^2{\bf r}
\nonumber\\
&+&\kappa^2\int[(|A_+|^4+|A_-|^4)/4-|A_+|^2|A_-|^2]d^2{\bf r},
\end{eqnarray}
where the dispersive operators $\hat D_\pm$ are
\begin{equation}
\hat D_\pm=\mp{i\partial_x}/({2\kappa})+{\partial_x^2}/({8\kappa^2})
-{\partial_y^2}/({4\kappa^2})+\cdots
\end{equation}

It should be emphasized that a (spatially non-uniform) linear Bragg coupling 
between the forward wave and the backward wave  introduces essentially new
effects in the dynamics, in comparison with the flat-bottom case.
It is already seen from the linear-wave dispersion relations corresponding 
to the simplest case $\varepsilon=const$ and $\Delta=const>0$. 
There are two branches in the spectrum (type-1 and type-2 waves), 
behaving quite non-trivially,

\begin{equation}
\tilde\omega^{-1}\Omega_{1,2}({\bf k})=-\varepsilon
\pm\sqrt{\Delta^2+\frac{k_x^2}{4\kappa^2}}+ 
\frac{k_y^2}{4\kappa^2}-\frac{k_x^2}{8\kappa^2}+\dots
\end{equation}
Various  nonlinear processes occur in this system, which create 
instabilities near corresponding resonant curves in ${\bf k}$-plane. 
For example, two type-1 waves with ${\bf k}_1={\bf 0}$ decay  into 
two type-2 waves with wave vectors $\pm{\bf k}_2$
($2\leftrightarrow 2$ process), if
\begin{equation}
2\Omega_1({\bf 0})\approx\Omega_2({\bf k}_2)+\Omega_2(-{\bf k}_2).
\end{equation}
Thus, a spatially homogeneous solution $A_+=A_-=A_0\exp(-i\Omega_0 t)$, 
with a small but finite constant amplitude $A_0$, is unstable. 
A standard linear analysis shows that a maximum of the instability increment 
is reached at the perpendicular direction, near 
${\bf k}_2=\pm{\bf e}_y \kappa\sqrt{8\Delta}$. 
Moreover, besides the above indicated instability, 
there is another, long-scale modulation 
instability of type-1 waves, which completely surrounds the wave vector 
${\bf k}=0$, unlike the situation for oppositely propagating waves at 
infinitely deep water (see, e.g.,  Refs.\cite{ORS2006,SKEMS2006}).
Therefore the tendency towards spontaneous formation of big waves is more 
strong if the bottom is (locally) periodic.
However, a more detailed analytical study of 
Eqs.(\ref{Aplus_2D_model_1Dbed}-\ref{Aminus_2D_model_1Dbed}), including
full stability analysis and particular solutions, will be a subject of future
research. In this work we only present results of numerical simulations 
which give a general  impression about the dynamics,
with particular attention to the phenomenon of wave collapses.

\begin{figure}
\begin{center}
\epsfig{file=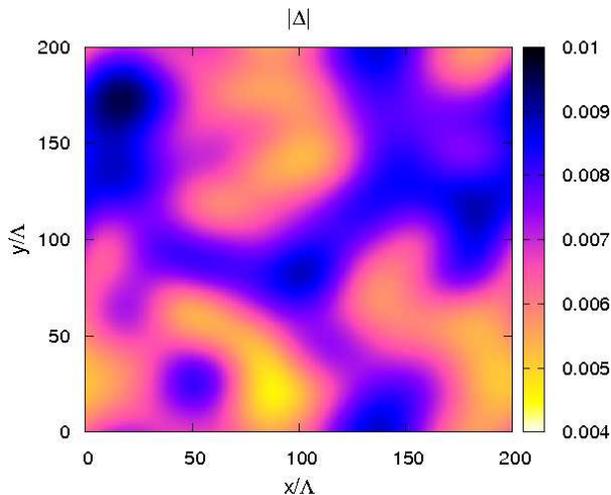,width=80mm}
\end{center}
\caption{(Color online). The absolute value of $\Delta(x,y)$.} 
\label{Delta} 
\end{figure}
\begin{figure}
\begin{center}
\epsfig{file=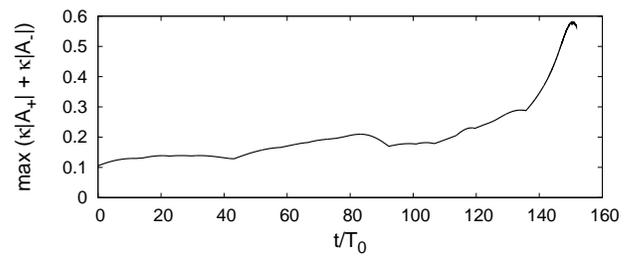,width=85mm}
\end{center}
\caption{The global maximum of wave amplitude versus time.} 
\label{A_max} 
\end{figure}
\begin{figure}
\begin{center}
\epsfig{file=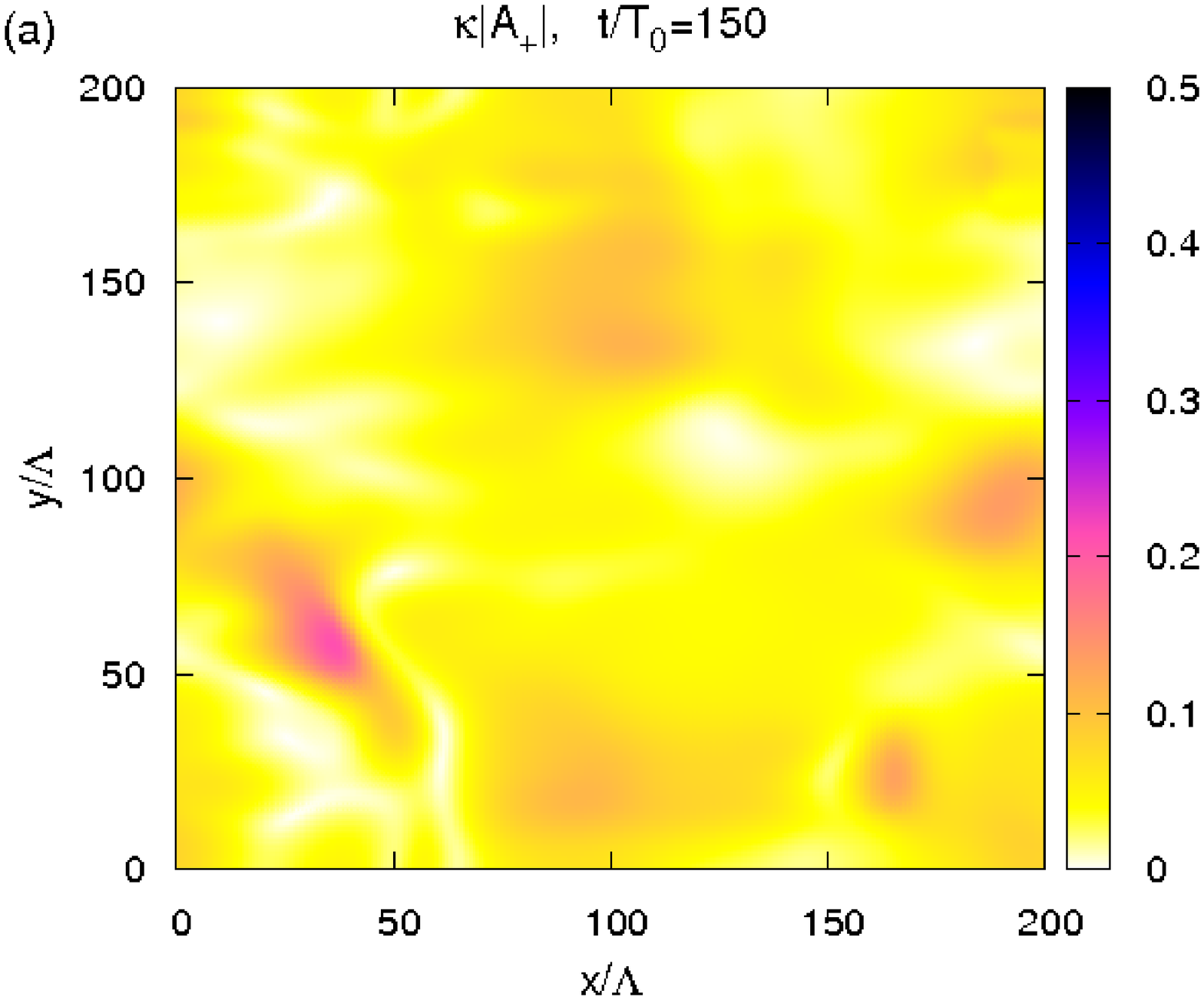,width=85mm}\\
\vspace{4mm}
\epsfig{file=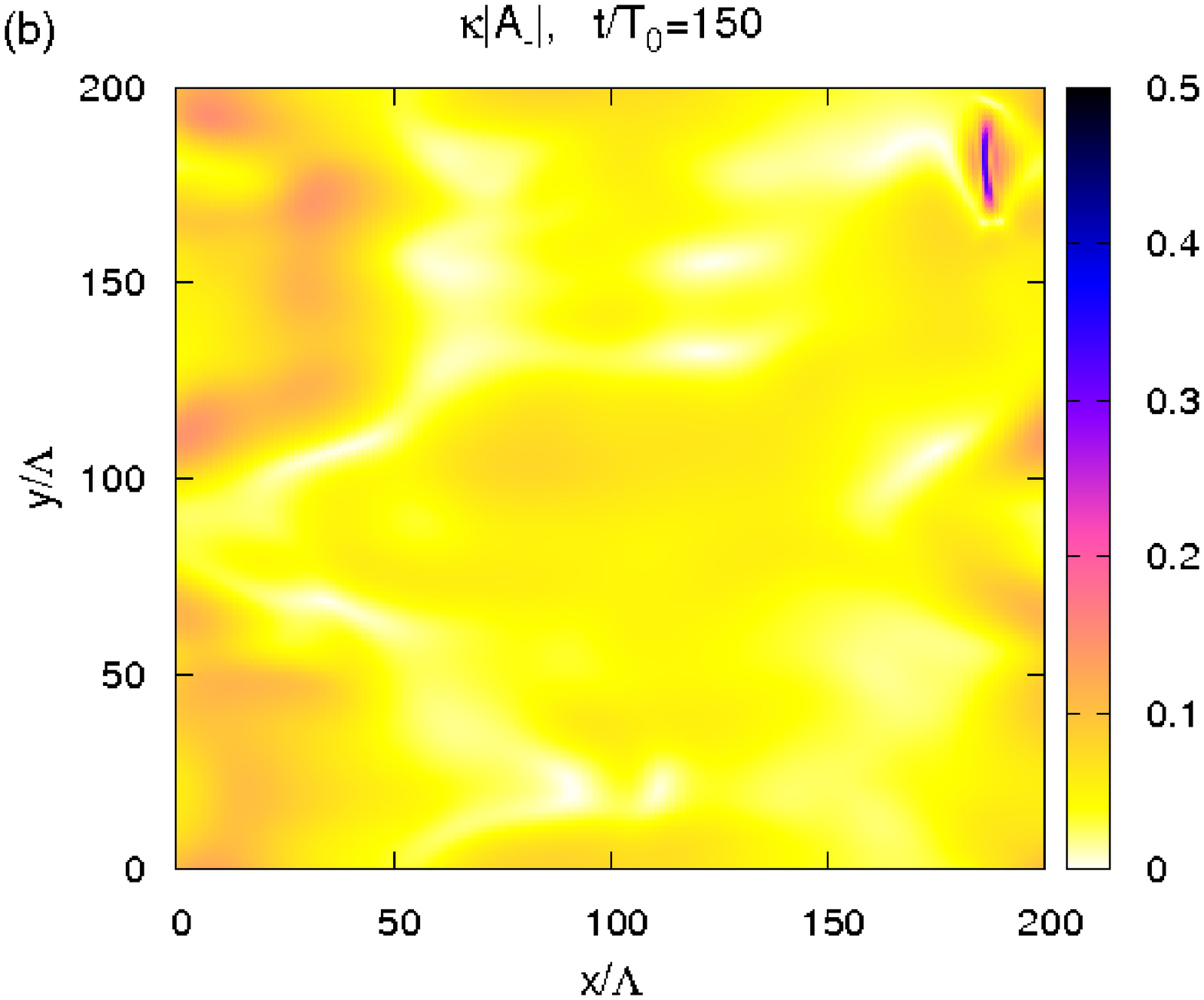,width=85mm}
\end{center}
\caption{(Color online). The absolute values $|A_\pm(x,y,t)|$ at $t/T_0=150$.
The localized energetic structure is seen near the right upper corner of
the $|A_-|$ map.} 
\label{t150} 
\end{figure}

The numerical simulations were performed in the standard dimensionless square 
domain $2\pi\times 2\pi$, with periodic boundary conditions and with
$\kappa=100$. In this example, function $\varepsilon$ was taken in the form
$\varepsilon=0.014( 1 + 0.1\cos x + 0.1\cos y )$, while
$\Delta(x,y)=0.007[1+0.05\xi(x,y)]$ was used, where a
complex function $\xi(x,y)$ contained  discrete 
Fourier harmonics within $-5\le k_x\le 5$ and $-5\le k_y\le 5$,
with amplitudes decaying as $\exp[-0.1(k_x^2+k_y^2)]$ and
with pseudo-random phases 
(see Fig.\ref{Delta}, where a map of the absolute value $|\Delta(x,y)|$ 
is shown). The initial conditions were $\kappa A_+=0.08+a_+(x,y)$ and 
$\kappa A_-=a_-(x,y)$, where functions $a_\pm(x,y)$ had the Fourier spectra
with  amplitudes behaving as $0.001\exp[-0.04(k_x^2+k_y^2)]$ and
with pseudo-random phases. Thus, the most part of wave energy was
initially concentrated in the forward wave. However, after few tens
of wave periods, the energy has been redistributed between the forward and
backward waves due to the inhomogeneous Bragg coupling. Wide regions
of slightly higher wave amplitude were formed, as a result of interplay
between random initial conditions and random $\Delta$, which process is
reflected in Fig.\ref{A_max}.  After that, self-focusing nonlinear
interactions came into play and rapidly produced the big wave 
which is seen in Fig.\ref{t150}. It is interesting to note that 
the localized structure was developed only in one of the two wave envelopes:
there is no clear sign of the same big wave in the opposite-wave amplitude.

The $x$-size of the developed high wave group 
contains just 1-2 wave lengths, while the $y$-size is larger, 
about 10 wave lengths. Strictly speaking, the assumption of narrow spectra
for $A_\pm$ is violated in this situation, so the system 
(\ref{Aplus_2D_model_1Dbed}-\ref{Aminus_2D_model_1Dbed}) can
describe the final stage of wave collapse only qualitatively.
To get indirect confirmations for the wave collapses, 
we compared the above approximate results to 
analogous simulations of full systems of the type (\ref{b_k_equation}), 
for some simple functions $T_m({\bf k}_1,{\bf k}_2;{\bf k}_3,{\bf k}_4)$
which possess the same property (\ref{T1212}) and thus mimic the true 
complicated matrix element $T({\bf k}_1,{\bf k}_2;{\bf k}_3,{\bf k}_4)$.
In particular, we considered 
\begin{eqnarray}
T_m\propto
|{\bf k}_1|^{\frac{1}{2}}|{\bf k}_2|^{\frac{1}{2}}
|{\bf k}_3|^{\frac{1}{2}}|{\bf k}_4|^{\frac{1}{2}}
(|{\bf k}_1+{\bf k}_2|+|{\bf k}_3+{\bf k}_4|&&\nonumber\\
-|{\bf k}_1-{\bf k}_3|-|{\bf k}_1-{\bf k}_4|
-|{\bf k}_2-{\bf k}_3|-|{\bf k}_2-{\bf k}_4|),&&
\end{eqnarray}
and there the wave collapses were observed as well (we do not discuss
here subtle points related to such systems). 
In the reality, however, higher-order nonlinearities 
become important at $|\kappa A_\pm|\gtrapprox 0.3$ and they produce
wave breaking manifested as the well-known ``white caps''.  
Therefore the question about maximal wave height at the finite stage 
of collapse can be fully answered only through real-world experiments.

To conclude, in the present work we have derived and then numerically 
simulated nonlinear  equations of motion for complex water-wave amplitudes 
coupled both linearly through the (spatially non-uniform) Bragg interaction, 
and non-linearly through the cross-modulation terms.
In the numerical experiments, we have observed spontaneous formation 
of highly energetic localized structures which look like wave
collapses rather than like solitons. The final stage of these 
wave collapses is actually beyond applicability of our weakly-nonlinear model.
That is why some real-world experiments or at least fully nonlinear simulations 
are desirable. Unfortunately, large-scale simulations in the framework of the 
exact Eulerian dynamics, for instance with the help of the existing boundary 
integral methods, are hardly possible at the moment because of their extremely 
time-demanding implementations. Perhaps, the weakly non-planar but fully 
nonlinear equations of motion for potential water waves over
quasi-one-dimensional topography, derived in Ref.\cite{R2005PRE},  
might be useful as an intermediate step towards accurate numerical results.

These investigations were supported by RFBR 
(grants 09-01-00631 and 07-01-92165),
by the ``Leading Scientific Schools of Russia'' grant 4887.2008.2,
and by the Program ``Fundamental Problems of Nonlinear Dynamics'' 
from the RAS Presidium.


\begin{thebibliography}{99}

\bibitem{CM1987} 
W. Chen and D.L. Mills, Phys. Rev. Lett. {\bf 58}, 160 (1987).

\bibitem{AW1989} A. B. Aceves and S. Wabnitz, 
Phys. Lett. A {\bf 141}, 37 (1989).

\bibitem{ESdSKS1996}
B. J. Eggleton, R. E. Slusher, C. M. de Sterke {\it et al.}, 
Phys. Rev. Lett. {\bf 76}, 1627 (1996). 

\bibitem{CJ1989} 
D.N. Christodoulides and R.I. Joseph, Phys. Rev. Lett. {\bf 62}, 1746 (1989).

\bibitem{PPLM1997} 
T. Peschel, U. Peschel, F. Lederer, and B. A. Malomed,
Phys. Rev. E {\bf 55}, 4730 (1997).

\bibitem{BPZ1998}
I.V. Barashenkov, D.E. Pelinovsky, and E.V. Zemlyanaya, 
Phys. Rev. Lett. {\bf 80}, 5117 (1998). 

\bibitem{RCT1998}
A. de Rossi, C. Conti, and S. Trillo, Phys. Rev. Lett. {\bf 81}, 85 (1998). 

\bibitem{CTA2000} C. Conti, S. Trillo, and G. Assanto,
Phys. Rev. Lett. {\bf 85}, 2502 (2000). 
 
\bibitem{IdS2000} 
T. Iizuka and C. Martijn de Sterke, Phys. Rev. E {\bf 61}, 4491 (2000).

\bibitem{CT2001} 
C. Conti and S. Trillo, Phys. Rev. E {\bf 64}, 036617 (2001).

\bibitem{CMMNSW2008} 
K. W. Chow, I. M. Merhasin, B. A. Malomed {\it et al.}, 
Phys. Rev. E {\bf 77}, 026602 (2008).

\bibitem{EC2003} N. Efremidis and D. N. Christodoulides,
Phys. Rev. A {\bf 67}, 063608 (2003).

\bibitem{PSK2004} D. E. Pelinovsky, A. A. Sukhorukov, and Yu. S. Kivshar,
Phys. Rev. E {\bf 70}, 036618 (2004).

\bibitem{MKTK2006} M. Matuszewski, W.  Krolikowski, M. Trippenbach,
and Y. S. Kivshar, Phys. Rev. A {\bf 73}, 063621 (2006).

\bibitem{R2008PRE-2} V. P. Ruban, Phys. Rev. E {\bf 77}, 055307(R) (2008).

\bibitem{R2008PRE-3} V. P. Ruban, Phys. Rev. E {\bf 78}, 066308 (2008).

\bibitem{R2004PRE} V. P. Ruban, Phys. Rev. E {\bf 70}, 066302 (2004).

\bibitem{Kharif-Pelinovsky} C. Kharif and E. Pelinovsky,
Eur. J. Mech. B/Fluids {\bf 22}, 603 (2003). 

\bibitem{RogueWaves2006} Special Issue:
Eur. J. Mech. B/Fluids {\bf 25}, 535-692 (2006).

\bibitem{MeiLi2004} C. C. Mei and Y. Li, Phys. Rev. E {\bf 70}, 016302 (2004).

\bibitem{GKN2007} J. Garnier, R. A. Kraenkel, and A. Nachbin, 
Phys. Rev. E {\bf 76}, 046311 (2007).

\bibitem{RA_Smith1998} R. A. Smith,  J. Fluid Mech. {\bf 363}, 333 (1998).

\bibitem{K1994} V. P. Krasitskii,  J. Fluid Mech. {\bf 272}, 1 (1994).

\bibitem{Z1999} V. Zakharov, Eur. J. Mech. B/Fluids {\bf 18}, 327 (1999).

\bibitem{ORS2006} M. Onorato, A. R. Osborne, and M. Serio,
Phys. Rev. Lett. {\bf 96}, 014503 (2006).

\bibitem{SKEMS2006} P. K. Shukla, I. Kourakis, B. Eliasson {\it et al.}, 
Phys. Rev. Lett. {\bf 97}, 094501 (2006). 

\bibitem{R2005PRE} V. P. Ruban, Phys. Rev. E {\bf 71}, 055303(R) (2005).

\end{thebibliography}
\end{document}